\documentclass[sigconf, review=false]{acmart}

\usepackage{lipsum}
\usepackage{amsfonts}
\usepackage{graphicx}
\usepackage{epstopdf}
\usepackage{algorithmic}
\usepackage{listings}
\usepackage{mathtools}
\usepackage{siunitx}
\usepackage{multirow}
\usepackage{booktabs} 
\usepackage{graphicx}
\usepackage{amsmath}
\usepackage{amssymb}
\usepackage{algorithm}
\usepackage{amsfonts}
\usepackage{mathtools}
\DeclarePairedDelimiter{\ceil}{\lceil}{\rceil}
\ifpdf
  \DeclareGraphicsExtensions{.eps,.pdf,.png,.jpg}
\else
  \DeclareGraphicsExtensions{.eps}
\fi

\lstdefinestyle{customc}{
  belowcaptionskip=1\baselineskip,
  breaklines=true,
  xleftmargin=\parindent,
  language=C,
  showstringspaces=false,
  basicstyle=\footnotesize\ttfamily,
  keywordstyle=\bfseries\color{green!40!black},
  commentstyle=\itshape\color{purple!40!black},
  identifierstyle=\color{blue},
  stringstyle=\color{orange},
}
\acmConference[]{May'19}
\acmYear{2019}
\copyrightyear{2019}

\begin{document}
\title{A Flexible Framework for Parallel Multi-Dimensional DFTs}

\author{Doru Thom Popovici}
\affiliation{%
  \institution{Carnegie Mellon University}
}
\email{dpopovic@alumni.cmu.edu}

\author{Martin D. Schatz}
\affiliation{%
  \institution{Facebook}
  }
\email{mschatz@fb.com}

\author{Franz Franchetti\\ Tze Meng Low}
\affiliation{%
  \institution{Carnegie Mellon University}
}
\email{{franzf, lowt}@cmu.edu}

\renewcommand{\shortauthors}{D. T. Popovici et al.}

\begin{abstract}
Multi-dimensional discrete Fourier transforms (DFT) are typically decomposed into multiple 1D transforms. Hence, parallel implementations of any multi-dimensional DFT focus on parallelizing within or across the 1D DFT. Existing DFT packages exploit the inherent parallelism across the 1D DFTs and offer rigid frameworks, that cannot be extended to incorporate both forms of parallelism and various data layouts to enable some of the parallelism. However, in the era of exascale, where systems have thousand of nodes and intricate network topologies, flexibility and parallel efficiency are key aspects all multi-dimensional DFT frameworks need to have in order to map and scale the computation appropriately. In this work, we present a flexible framework, built on the Redistribution Operations and Tensor Expressions (ROTE) framework, that facilitates the development of a family of parallel multi-dimensional DFT algorithms by 1) unifying the two parallelization schemes within a single framework, 2) exploiting the two different parallelization schemes to different degrees and 3) using different data layouts to distribute the data across the compute nodes. We show the need of a versatile framework and thus a need for a family of parallel multi-dimensional DFT algorithms on the K-Computer, where we show almost linear strong scaling results for problem sizes of $1024^3$ on $32k$ compute nodes.

\end{abstract}

%
%

\ccsdesc[500]{Theory of computation~Parallel algorithms}
\ccsdesc[500]{Software and its engineering~Application specific development environments}
\ccsdesc[300]{Computing methodologies~Parallel algorithms}

\keywords{3D FFTs, Distributed Systems, Performance, Scalability}

\maketitle

\section{Introduction}

The multi-dimensional discrete Fourier transform (DFT) has proven to be an ubiquitous mathematical kernel, that is widely used in a multitude of applications from different scientific fields like molecular dynamics~\cite{Plimpton97particle-meshewald, sirk2013characteristics, plimpton2012computational}, material sciences~\cite{chang2004numerical, vay2018warp, almgren2013nyx, lebensohn2012elasto}, quantum mechanics~\cite{kendall2000high,valiev2010nwchem,straatsma2011advances, canning2000parallel}. As most of these applications spend a significant amount of the total execution on computing multi-dimensional DFTs, it is vital that parallel multi-dimensional DFT be efficient as we move into the exa-scale era. As multi-dimensional DFTs are defined in terms of multiple 1D DFTs and, parallel implementations of the multi-dimensional DFTs can be classified into two distinct classes. The first class focuses on exploiting parallelism within the 1D DFTs, while the second class exploits the inherent parallelism across multiple distinct 1D DFTs. Most state-of-the-art frameworks for computing 3D DFTs like FFTW~\cite{FFTW}, P3DFFT~\cite{pekurovsky_p3dfft:_2012-1}, FFTE~\cite{DBLP:conf/ppam/Takahashi09} opt to parallelize across the 1D DFTs. 

\begin{figure}[t]
    \centering
    \includegraphics[width=0.45\textwidth]{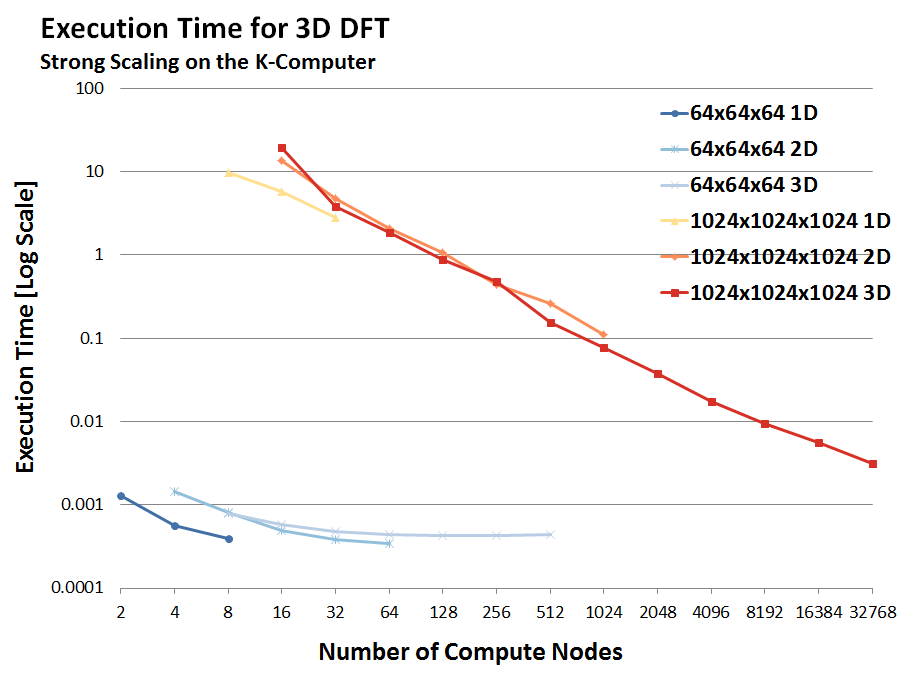}
    \caption{Strong scaling results for two 3D DFTs of size $64^3$ and $1024^3$ using the slab-pencil, pencil-pencil-pencil and volumetric decomposition on the K-Computer. The execution time is shown in log scale.}
    \label{fig:1024strongk}
\end{figure}

We illustrate the limitations of the conventional approach of parallelizing multi-dimensional DFTs in Figure~\ref{fig:1024strongk}, where we report the performance of three different parallelization schemes for computing the 3D DFT of $64^3$ and $1024^3$ on the K-Computer. Notice that for data set of size $64^3$, parallelizing the 3D-DFT on a 1D grid of processor such that each processors computes a 2D plane yielded the shortest execution time. However, for the $1024^3$ input, the same parallelization scheme scales only to $32$ processors, while better performance can be attained by switching to a parallelization scheme comprising of a 2D grid of processors where each processor computes a small batch of 1D DFTs. Just as with the 1D parallelization scheme, the parallelism with a 2D grid of processors is restricted to at most $1024$ processors. By parallelizing within the 1D DFT for the last dimension, we obtain an even shorter execution time using a much greater number of processors. These observations highlight two major limitations of existing approaches: 1) there is a need for a framework that allows one to seamlessly switch between parallel algorithms for computing the multi-dimensional DFT as different parallel algorithms are superior for different problem sizes, and 2) there is a need to parallelize both across and within the 1D DFTs to scale to larger number of compute nodes.

There is an increasing interest in scaling the parallel implementations of the multi-dimensional DFT to higher number of nodes using higher dimensional grids as shown in Jung et. al.~\cite{jung2016parallel}. These newer algorithms are a subset of the parallel DFT algorithms highlighted in Johnson et. al.~\cite{johnson2003recursive}, that showed that a significant number of algorithms can arise from the parallelization both within and across multiple 1D DFTs. However, these algorithms are largely unexplored in practice as the existing parallel DFT frameworks make implementing them quite difficult. Due to the rigidity of the existing frameworks, the resulting performance is still unsatisfactory. Therefore, in this work we present a flexible framework for employing different parallelization schemes to compute the multi-dimensional DFT on a multi-dimensional computation grid of processors. By recognizing that the inputs and outputs of a multi-dimensional DFT are multi-dimensional arrays of data, we leverage insights from the multi-linear algebra (tensor) community to simplify the management of communication and data layout on a multi-dimensional grid. 

{\bf{Contributions.}} The paper makes the following contributions:
\begin{itemize}
    \item A flexible framework that unifies both parallelization schemes for multi-dimensional DFTs on multi-dimensional grids, built on top a tensor framework.
    \item Demonstrate the need for a family of parallel DFT algorithms in order to attain high performance for any given problem size and network configurations.
\end{itemize}

\section{The Discrete Fourier Transform}

In this section, we briefly present the decomposition of both the 1D and multi dimensional DFTs, expressing the algorithms in terms of linear algebra operations and outlining the possibilities for parallelization.

\begin{figure}[t]
    \centering
    \includegraphics[width=0.45\textwidth]{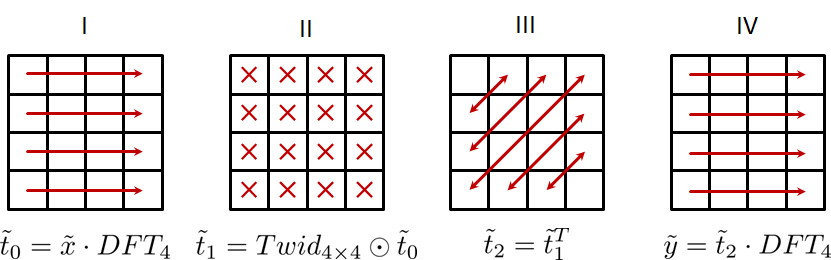}
    \caption{The decomposition of the 1D DFT using the Cooley-Tukey algorithm for a given problem size of $16 = 4\cdot 4$. The algorithm requires four steps, namely two batch DFTs (I) and (IV) of size $4$, a point-wise operation (II) and a transposition (III). Data is stored in column major order.}
    \label{fig:1dct}
    \vspace{-5mm}
\end{figure}

\subsection{The 1D Discrete Fourier Transform}

The 1D DFT is a matrix-vector multiplication, where given the input $x$, the output $y$ is obtained as
\begin{align}\label{eq:1ddft}
    y = DFT_n\cdot x.
\end{align}
The $DFT_n$ is the $n \times n$ DFT dense matrix, defined as 
\begin{align}
    DFT_n = \begin{bmatrix}\omega^{l k}_n\end{bmatrix}_{0\leq l, k < n},
\end{align}
with $\omega_n = e^{-k\frac{2\pi}{n}}$ being the complex root of unity. Typically, the computation of the DFT is implemented using the Fast Fourier Transforms (FFT), where instead of performing $O(n^2)$ complex arithmetic operations by doing the matrix-vector multiplication, a recursive decomposition of the DFT matrix is performed to obtain an $O(n \log(n))$ algorithm. The most widely-know of these algorithms is the Cooley-Tukey algorithm~\cite{cooley1965algorithm}. The Cooley-Tukey algorithm is described as a factorization of the DFT matrix of size $n$ when $n$ is a composite number such as $n = n_0 n_1$.

\begin{figure}[b]
    \centering
    \includegraphics[width=0.4\textwidth]{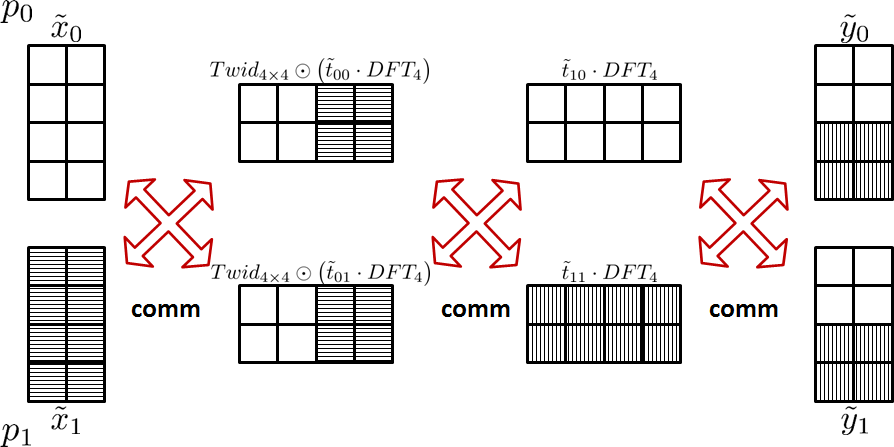}
    \caption{The parallel implementation of the Cooley-Tukey algorithm for the $DFT_{16}$ on $p = 2$ processors. Data is distributed between the processors, each processor does its local computation and then exchanges the data. This parallel implementation requires three communication stages.}
    \label{fig:parallelct}
\end{figure}

An alternative view of the mathematical description of the Cooley-Tukey algorithm defined in~\cite{cooley1965algorithm} is as follows:
\begin{align}\label{eq:ct}
    \tilde{y} = \left(Twid_{n_0\times n_1}\odot\left(\tilde{x}\cdot DFT_{n_1}\right)\right)^T\cdot DFT_{n_0},
\end{align}
where $\tilde{x}$ and $\tilde{y}$ are two matrices of size $n_0\times n_1$ and $n_1\times n_0$, that represent the input $x$ and output $y$, respectively. The columns of matrix $\tilde{x}$ ($\tilde{y}$) are $n_1$ ($n_0$) contiguous sub-vectors (i.e. $x_i$, $0\leq i < n_0$ or $y_i$, $0\leq i < n_1$) of the input (output) vector $x$ ($y$), and are obtained by splitting $x$ ($y$) into sub-vectors of size $n_0$ ($n_1$) as follows
\begin{align*}
    \tilde{x} = \begin{bmatrix}x_0 & | & x_1 & |& \ldots&|& x_{n_1-1}\end{bmatrix}.
\end{align*}
The matrix $\tilde{y}$ is similarly constructed.

The reason for the formulation of the Cooley-Tukey algorithm as described by Equation~\ref{eq:ct} is that the four stages of the algorithm (as depicted in Figure~\ref{fig:1dct}) are made explicit. The first step applies the $DFT_{n_1}$ on the rows of the matrix $\tilde{x}$ (I), assuming data is stored in column major order. The elements within each column of either matrix $\tilde{x}$ or $\tilde{y}$ are in consecutive locations in memory. This means that in order to compute the first stage of the DFT, elements are accessed at a stride of $n_0$. The result of the first compute stage is then point-wise multiplied with the so-called twiddle matrix (II), which is defined as
\begin{align}
    Twid_{n_0\times n_1} = \begin{bmatrix}\omega^{k l}_n\end{bmatrix}_{0\leq l < n_0\text{ and } 0\leq k < n_1},
\end{align}
where $\omega_n$ represents the complex roots of unity. The resultant matrix is then transposed (III) and finally the second DFT of size $n_0$ is applied in the rows (IV). Again, the elements required for the fourth stage of the computation are read at a stride. The second DFT cannot start computation until all previous stages have been completed, because of the transposition stage.

\subsection{Parallelizing the 1D DFT}

Parallelizing the 1D DFT on $p$ processors requires the parallelization of all four compute stages. Traditionally, the input matrix $\tilde{x}$ (stored in column major order) is split such that each processor receives $n_1 / p$ columns of size $n_0$. Distributing the data using this data layout implies that computation cannot start since each processor does not have the necessary data points. As such, an all-to-all communication step is needed to re-distributes the data such that each processor receives $n_0/p$ rows of size $n_1$. The first and second stage of the Cooley-Tukey algorithm can then be applied locally. A second all-to-all communication is performed to transpose (stage III) the global data across the different processors. After this communication step, each processor owns $n_1 / p$ rows of size $n_0$ and thus can locally compute the last stage of the algorithm. Storing the data in the correct order requires a third communication step. The described computation gives the so-called six step algorithm~\cite{VanLoan} defined as 
\begin{align}
    \tilde{y} = \left(DFT_{n_0}\cdot\left(Twid_{n_0\times n_1}\odot\left(DFT_{n_1}\cdot\tilde{x}^T\right)\right)^T\right)^T.
    \label{eqn:ctparallel}
\end{align}
The parallel implementation of the six step algorithm is depicted in Figure~\ref{fig:parallelct}, and it requires a total of three communication steps. The astute reader will recognize that \emph{some of the communication steps can be avoided by storing the initial data in a different layout. This has been observed in literature, but seldom exploited in practice}.

\subsection{$n$-Dimensional DFTs}

\begin{figure}
    \centering
    \includegraphics[width=0.4\textwidth]{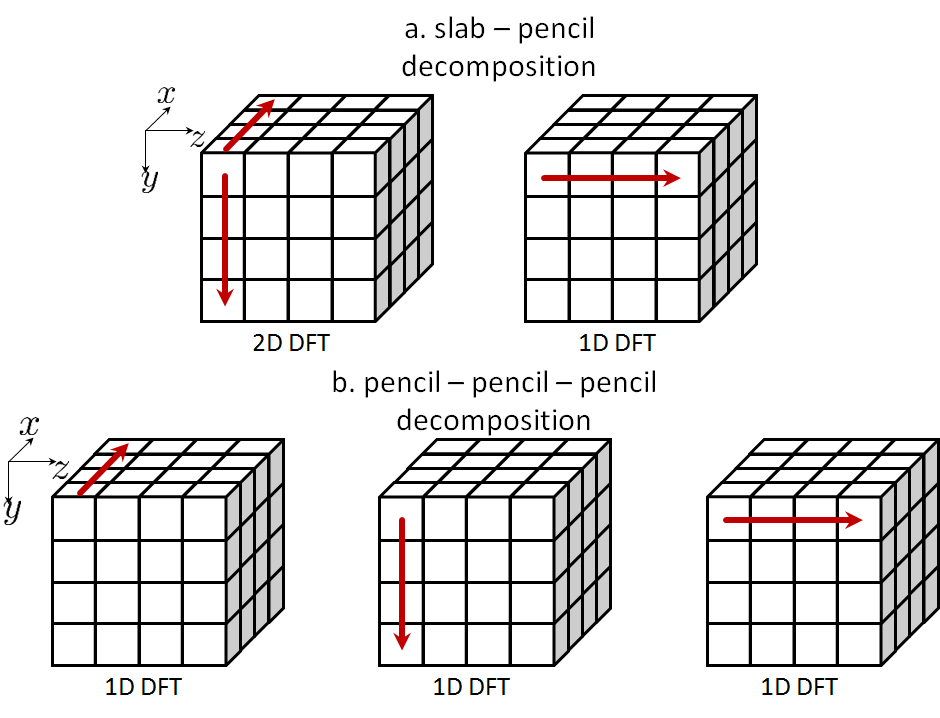}
    \caption{Two algorithms for computing the 3D DFT. The slab-pencil algorithm (a) decomposes the 3D DFT into a 2D DFT and a 1D DFT. When applied, the 2D DFT is decomposed into the corresponding 1D transforms. The pencil-pencil-pencil algorithm (b) decomposes the 3D DFT into three 1D DFTs applied in the corresponding dimensions.}
    \label{fig:3ddftopt}
    \vspace{-3mm}
\end{figure}

Multi-dimensional DFTs can be defined in terms of multiple 1D DFTs and multi-dimensional DFTs of lower dimensions. For example, Figure~\ref{fig:3ddftopt} shows two variants for computing the 3D DFT. The first algorithm represents the so-called slab pencil decomposition~\cite{swarztrauber1987multiprocessor, foster1997parallel, sweet1991ffts}, where the 3D DFT is decomposed into a 2D DFT followed by a batch of multiple 1D DFTs. The second algorithm represents the pencil-pencil-pencil decomposition~\cite{swarztrauber1987multiprocessor, foster1997parallel, sweet1991ffts}, where the 3D DFT is decomposed into three batches of 1D DFTs, where each 1D DFT is applied in the three dimensions.

The slab-pencil decomposition views the input (output) column vectors $x$ ($y$) as 2D matrices $\tilde{x}$ ($\tilde{y}$) of size $\left(n_0 n_1\right)\times n_2$. Mathematically the decomposition is expressed as
\begin{align}
    \tilde{y} = \left(DFT_{n_0\times n_1}\cdot\tilde{x}\right)\cdot DFT_{n_2}
\end{align}
Data is stored in column major, hence the 2D DFT is applied on the columns, while the 1D DFT is applied on the rows. 

The pencil-pencil-pencil algorithm views the input (output) vectors are reshaped into 3D cubes $\hat{x}$ ($\hat{y}$) of size $n_0\times n_1\times n_2$. The input (output) column vector $x$ ($y$) is decomposed into $n_2$ groups of $n_1$ subgroups of size $n_0$. Hence, the 3D cube $\hat{x}$ can be viewed as matrix of matrices such as
\begin{align}
    \hat{x} = \begin{bmatrix}\tilde{x}_0| & \tilde{x}_1| & \ldots| & \tilde{x}_{n_2}\end{bmatrix},
\end{align}
where each $\tilde{x}_i$ is the 2D matrix of size $n_0\times n_1$ for all values $0\leq i < n_2$. Mathematically, the pencil-pencil-pencil algorithm is expressed as
\begin{align}
    \hat{y} = \begin{bmatrix}\left(DFT_{n_0}\cdot\tilde{x}_0\right)\cdot DFT_{n_1}| & \ldots| & \left(DFT_{n_0}\cdot\tilde{x}_{n_2}\right)\cdot DFT_{n_1}\end{bmatrix}\cdot DFT_{n_2}
\end{align}
where the $DFT_{n_0}$ is applied in the depth dimension, the $DFT_{n_1}$ is applied in the column dimension and finally the $DFT_{n_2}$ is applied in the row dimension. Data is store in column major and therefore the dimension corresponding to $n_0$ is laid out in the fastest dimension in memory, while the dimension corresponding to $n_2$ is in the slowest dimension in memory. 

\begin{figure}[h]
   \centering
    \includegraphics[width=0.45\textwidth]{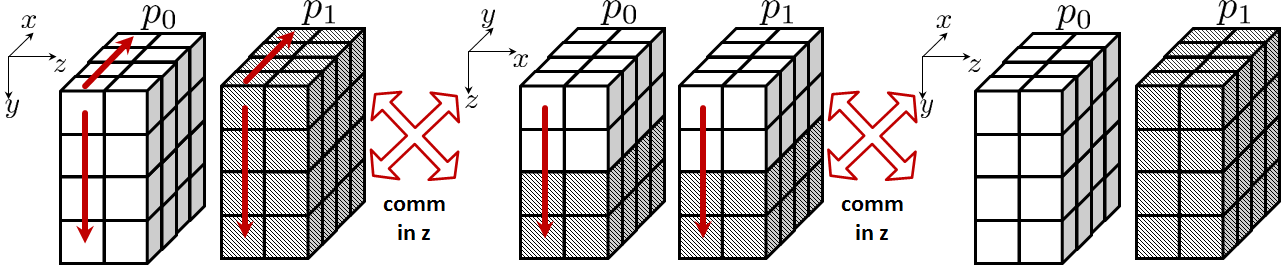}
    \caption{The parallel implementation of the slab-pencil algorithm for computing the 3D DFT on a 1D mesh of two processors. Each processor applies a 2D DFT followed by a 1D. The implementation requires two communications stages.}
    \label{fig:3dslabpencil}
\end{figure}

\begin{figure*}[t]
    \centering
    \includegraphics[width=\textwidth]{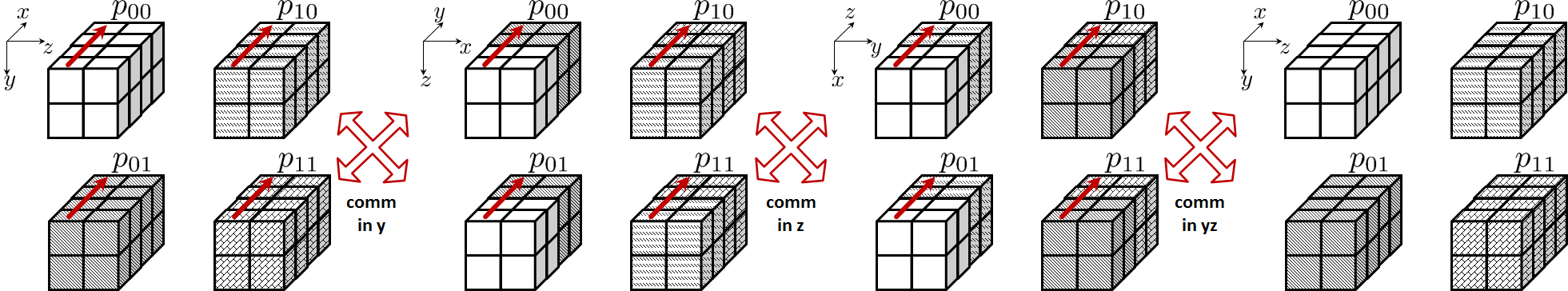}
    \caption{The parallel implementation of the pencil-pencil-pencil algorithm for computing the 3D DFTs on a 2D mesh of size $2\times 2$ processors. Each processor 1D DFTs in each dimension. The implementation requires three communications stages.}
    \label{fig:3dpencilpencil}
\end{figure*}

\subsection{Parallelizing the $n$-Dimensional DFTs}
Since the multi-dimensional DFT can be decomposed into multiple lower-dimensional DFTs, most prevalent DFT frameworks exploit this feature for parallelization. We illustrate this parallelization scheme on the two algorithms for computing the 3D DFT that were presented previously. Parallelizing the slab-pencil algorithm is performed by viewing the processors as a 1D grid and distributing the data cube such that each processor receives a slab of $n_2/p$ 2D data of size $n_0\times n_1$. Each processor then locally computes multiple 2D DFTs on the slabs. Data is then exchanged across processors so that each processor can compute $(n_0 \times n_1) / p$ 1D DFT of size $n_2$. Finally, another communication step is required to store the data in the correct format as seen in Figure~\ref{fig:3dslabpencil}. 

The pencil-pencil-pencil algorithm is parallelized in the two dimensions corresponding to $n_1$ and $n_2$ across a mesh of processors of size $p_y\times p_z$. Each processor receives a batch of 1D pencils as seen in Figure~\ref{fig:3dpencilpencil}. Each processor can apply the local 1D DFTs. In order to apply the other two 1D DFTs data is first exchanged between the processors in the $p_y$ dimension of the mesh, followed by the communication between the processors on the $p_z$ dimension. Each communication stage rotates the data cube from $xyz$ to $yzx$ and $zxy$ as seen in Figure~\ref{fig:3dpencilpencil}. Each communication step requires data to be packed and un-packed before and after the all-to-all communication. Finally, in order to store the data in the correct format a global communication between all processors is required. However, the DFT is typically used in larger applications like convolutions or PDE solvers, therefore the last stage of communication for both the slab-pencil and pencil-pencil-pencil can be dropped. This improves reduces execution time in favor of having the data in shuffled format.

\subsection{Limitations of existing parallelization schemes}

Notice that in both cases, the amount of parallelism is limited by the size of a particular dimension of the data. For the slab-pencil algorithm, the maximum number of processors that can participate in the computation is $\max({n_0, n_1, n_2})$ as each processor is assigned a 2D slab of data. Similarly for the pencil-pencil-pencil case, the number of participating processors is upper bound by $\max(n_0n_1, n_1n_2, n_0n_2)$ since each processor is assigned at least one 1D DFT to compute locally. The level of parallelism within the communication is also limited by the size of a particular dimension of the data. The slab-pencil distribution spreads the data on a one dimensional grid. Hence, exchanging the data for computing the 1D DFT require all processors to communicate. The pencil-pencil-pencil case parallelizes the communication in two dimensions, which is done between groups of processors. All frameworks that implement the parallelization across the 1D DFT depend on efficient all-to-all communications~\cite{bruck1997efficient, prisacari2013bandwidth}. However, as we will briefly show in the result section, the all-to-all communication cannot efficiently scale as the number of processors increase.

\section{$n$-Dimensional DFTs in ROTE}

The inputs and outputs of any multi-dimensional DFT are multi-dimensional data arrays. Multi-dimensional arrays are commonly referred to as tensors in the multi-linear algebra community. As such, we leverage the insights from that community for distributed data management on a distributed grid through the use of the Redistribution Operations and Tensor Expressions (ROTE) framework~\cite{schatz2015distributed}.

\subsection{The Redistribution Operations and Tensor Expressions Framework}

The Redistribution Operations and Tensor Expressions (ROTE) framework provides a formal notation for describing the data layout of tensors that are distributed on multi-dimensional meshes. By describing the data layouts before and after a data redistribution operation, 
ROTE provides the infrastructure to map the necessary data movement into a sequence of one or more collective communication subroutines to achieve the desired data movement. In order to make the paper self-contain, we present a high-level overview of the notation of ROTE and refer the reader to existing documentations on ROTE~\cite{schatz2015distributed} for more details.

\subsubsection{Describing Data and Processing Grid} Within ROTE, data and processing grids are viewed as multi-dimensional arrays or tensors. A tensor is a $d$-dimensional, or a $d$-mode, array. The order of a tensor represents the number of dimensions or the number of modes represented by the tensor. The dimension of each mode refers to the length or size of a specific mode. For example, the $2 \times 8$ matrix defined as
\begin{align}
    A^{\iota \eta} = 
    \begin{bmatrix}
        a_{0,0} & a_{0,1} & a_{0,2} & a_{0,3} & a_{0,4} & a_{0,5} & a_{0,6} & a_{0,7}\\
        a_{1,1} & a_{1,1} & a_{1,2} & a_{1,3} & a_{1,4} & a_{1,5} & a_{1,6} & a_{1,7}\\
    \end{bmatrix}
\end{align}
is an order-2 tensor. For consistency within the paper, we use the terms "dimensions", and "size" to refer to the order of the tensor, and dimension of the mode respectively.  To make things clearer, we annotate the matrix $A$ with two superscripts ($\iota$ and $\eta$) to denote the size of
the two modes such that $A$ is $\iota \times \eta$.

\begin{figure}
    \centering
    \includegraphics[width=0.45\textwidth]{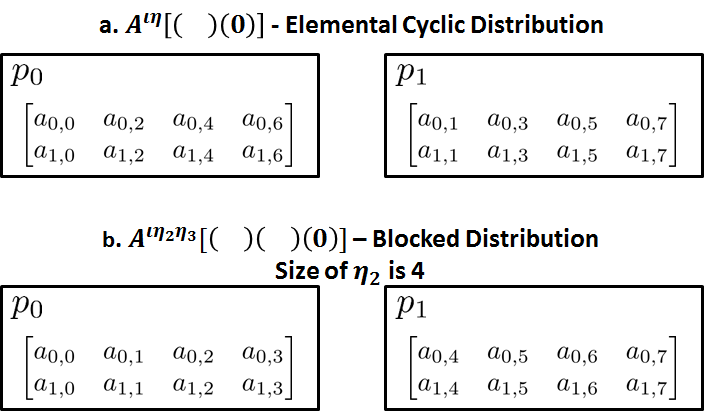}
    \caption{Data layouts for distributing the 2D matrix $A^{\iota\eta}$ on the 2D grid $G^{\rho\nu}$. A. represents the elemental cyclic distribution of the row dimension across the two processors. B. represents the block distribution for the row dimension.}
    \label{fig:datadistrib}
\end{figure}

Multi-dimensional processing grids are defined within ROTE as multi-dimensional arrays of processing units. Let $G$ be the d-dimensional grid size. Similar to the data tensors, an order-d $G$ grid will have $d$ dimensions and each dimension have a specific number of processors. For example, the 1D grid defined as
\begin{align}
    G^{\rho\nu} = 
    \begin{bmatrix} 
        p_{0} & p_{1}
    \end{bmatrix}
\end{align}
is a $1\times 2$ processor mesh on which data can be distributed. $\rho$ represents the column dimension of size one, while $\nu$ represents the row dimension of size two.

\subsubsection{Describing Global Data Distributions} One of ROTE's key aspects is the way in multi-dimensional data sets are described and distributed across multi-dimensional meshes. The description is based on the idea of distributing each dimension of the data over specific dimensions of the processor grid. We illustrate this aspect in the following example. Consider distributing the two dimensional tensor $A^{\iota \eta}$ across the 1D processor grid, $G^{\rho \nu}$ such that columns of $A^{\iota \eta}$ are distributed in a cyclic fashion across the processors (i.e. Figure~\ref{fig:datadistrib}). This is described in ROTE with as $A^{\iota \eta}$ having the distribution
\[
    A^{\iota\eta}[()(0)].
\]
To understand the expression, note that there are two pairs of round parenthesis within the square brackets. The pair of round brackets tells us that $A^{\iota\eta}$ is a two-dimensional tensor. In addition, note that there is only a single value, $0$, within the round brackets. This number specifies that the specific dimension of the data is distributed on the dimension of the processor grid specified by that value. Therefore, the zero within the second parenthesis tells us that the second dimension of $A$, i.e. the columns of $A^{\iota\eta}$, is distributed across the $0$-th dimension of the processor grid. The distribution is an elemental-cyclic distribution of the columns of $A^{\iota\eta}$ across the grid.

The notation can easily be extended to describe blocked data layouts. Blocked distributions are also an important data layout for the computation of the multi-dimensional DFT. The key difference between the blocked and the elemental-cyclic distributions is the number of consecutive elements that are distributed to each of the processors on the mesh. The elemental-cyclic distribution assumes the data being distributed at the granularity of one element, while the blocked distribution assumes that each processor receives $b$ consecutive elements. Expressing the blocked distribution of the matrix $A$ in ROTE notation can be done as follows
\begin{align}
    A^{\iota\eta_0\eta_1}[()()(0)],
\end{align}
where the column dimension of size $\eta = \eta_0 \eta_1$ is blocked or tiled by the $\eta_0$ size.

Blocking or tiling the data with a block of size $b$ can be interpreted as adding an another dimension to the data set as shown in~\cite{bondhugula2008practical}. In the previous example, the dimension described by $\eta$ is split into two, namely $\eta_0$ and $\eta_1$ and $\eta_0$ is set to the block size (in this case, 4).  The dimension corresponding to $\eta_1$ is distributed across the two processors on the grid as denoted by the zero value within the third pair of parenthesis. Each processor receives a block of $\eta_0$ elements in a cyclic fashion.

\begin{figure}
    \centering
    \includegraphics[width=0.45\textwidth]{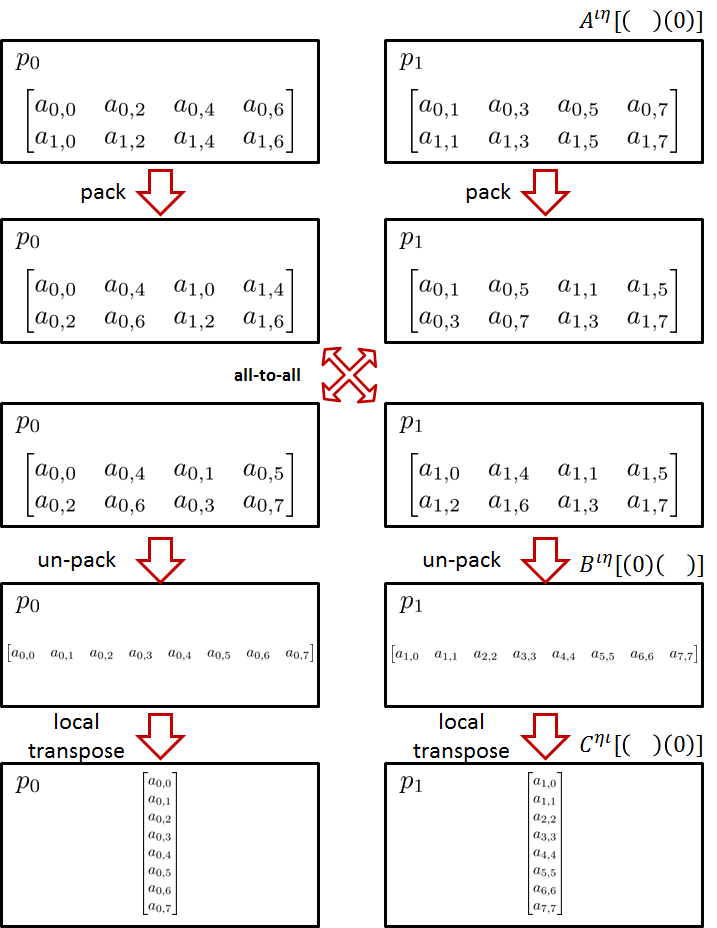}
    \caption{Transposing the matrix $A^{\iota\eta}[()(0)]$. The transposition is broken into a block transposition and multiple within-block transpositions. The operations from matrix $A^{\iota\eta}[()(0)]$ to matrix $B^{\iota\eta}[(0)()]$ represents the block transposition over the network (packing, all-to-all, un-packing). The last stage $C^{\eta\iota}[()(0)]$ does the local transpositions.}
    \label{fig:dataredist}
\end{figure}

\subsubsection{From Distribution to Communication}
ROTE implements data communication over the computation grid by specifying the original and final data layout. For example, the following expression
\begin{align}
    B^{\iota\eta}[(0)()] = A^{\iota\eta}[()(0)] 
\end{align}
describes the original tensor $A$, where the columns of the matrix $A$ are distributed element cyclic across the processor grid, being mapped to the new tensor $B$, where the rows of the tensor are distributed in an elemental-cyclic fashion. ROTE maps the above description to the all-to-all collective communication. Given the before and after data distributions, the infrastructure within ROTE determines the necessary packing, collective communication and un-packing required. The detailed steps that ROTE takes in order to perform the redistribution of the tensor $A$ from the initial distribution to the final distribution expressed as tensor $B$ are captured in Figure~\ref{fig:dataredist}.

\begin{figure*}[t]
    \centering
    \includegraphics[width=\textwidth]{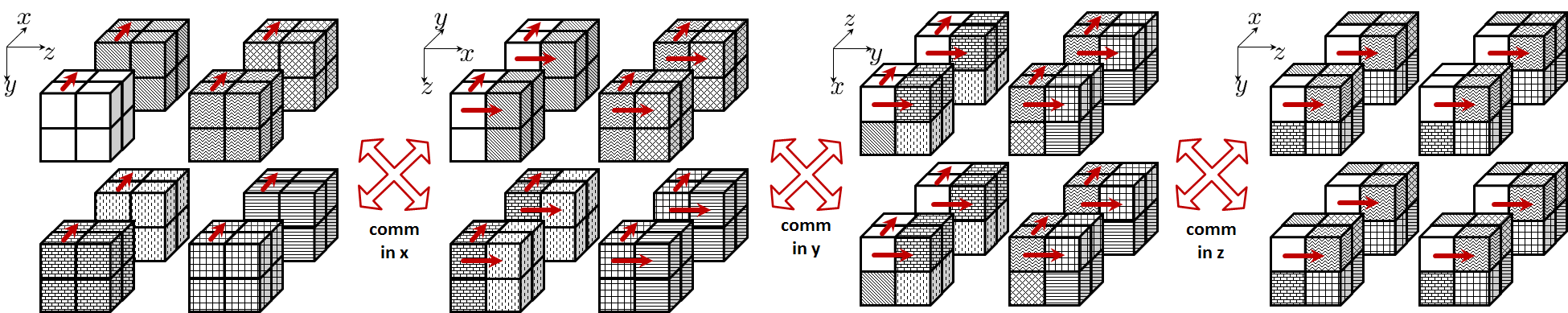}
    \caption{The parallel implementation of the 3D DFT using the volumetric decomposition. The data is distributed elemental cyclic in all three dimensions across a 3D mesh of processors of size $2\times 2\times 2$. The implementation requires three communication steps and the elemental cyclic distribution is preserved between the computed stages.}
    \label{fig:cuboid}
\end{figure*}

\subsection{Parallel 1D DFT using ROTE}
In this section, we describe how ROTE can be used as our framework for computing the 1D DFT using the Cooley-Tukey algorithm. Recall that 
the Cooley-Tukey algorithm decompose the computation of the 1D DFT of size $n = n_0 n_1$ into four stages involving computations with two dimensional tensors (i.e. matrices).  
Using the notation from ROTE, we can re-express Equation~\ref{eq:ct} as
\begin{align}\label{eq:cteinstein}
    \tilde{Y}^{\eta_1\eta_0} = \left(\left(Twid^{\iota_0\eta_1}_{n_0\times n_1}\odot\left( \tilde{X}^{\iota_0\iota_1}\cdot DFT^{\iota_1\eta_1}_{n_1}\right)\right)^T\right)\cdot DFT^{\iota_0\eta_0}_{n_0}.
\end{align}
${\iota_0\iota_1}$ and ${\eta_1\eta_0}$ represent the size of the two-dimensional tensors of the input and output respectively.

In the following paragraphs, we derive the 1D DFT parallelization across $p$ processors using ROTE's notation. Recall that the input matrix is distributed across the processors (in a one dimensional grid) such that each processor has $\eta_1/p$ columns. However under this data layout, the computation requires an initial redistribution of the input data such that each processor holds the row of the input matrix $\tilde{X}^{\iota_0\iota_1}$. Naturally, this suggest that the input data (to avoid the redistribution at the start) has to be distributed as
\[
\tilde{X}^{\iota_0\iota_1}[(0)()],
\]
where the rows are distributed in elemental cyclic fashion. Notice that while $\iota_0 \times \iota_1$ must be equal to the input size, the Cooley-Tukey algorithm places no other constraints on what $\iota_0$ and $\iota_1$ must be. However, it is natural to make $\iota_1 = p$ since that will ensure that the rows are cyclic-ly distributed across all $p$ processors.

The computation stage of the parallel 6-steps algorithm is then able to perform the local DFT computation, followed by the local point-wise multiplication with the twiddle matrix $Twid_0$. Since the computation is an element-wise complex multiplication it follows that the distribution of the twiddle matrix $Twid_0$ must be the same with that of the input $\tilde{X}$, i.e.
\[
 Twid_0^{\iota_0\eta_1}[(0)()].
\]

After the first two steps, the intermediate result in the temporary array $T$ is as follows
\begin{align}
    T_0^{\iota_0\eta_1}[(0)()] = \left(Twid^{\iota_0\eta_1}_{n_0\times n_1}[(0)()]\odot\left( \tilde{X}^{\iota_0\iota_1}[(0)()]\cdot DFT^{\iota_1\eta_1}_{n_1}\right)\right),
\end{align}
where $T_0^{\iota_0\eta_1}[(0)()]$ has the same distribution as the input $\tilde{X}^{\iota_0\iota_1}$.

The 1D DFT requires a transposition of the array $T_0^{\iota_0\eta_1}$. Figure~\ref{fig:dataredist} shows the steps required to perform the transposition. First the block transposition is applied such that
\begin{align}
    T_1^{\iota_0\eta_1} [()(0)] = T_0^{\iota_0\eta_1}[(0)()].
\end{align}
This operation is mapped to the all-to-all collective communication. Each processor then transposes the data locally such that 
\begin{align}
    T_2^{\eta_1\iota_0}[(0)()] &= T_1^{\iota_0\eta_1} [()(0)]
\end{align}
Finally, each processor applies its local computation of the last stage of the Cooley Tukey algorithm. The output $Y$ is computed as
\begin{align}
    \tilde{Y}^{\eta_1\eta_0}[(0)()] &= T_2^{\eta_1\iota_0}[(0)()]\cdot DFT^{\iota_0\eta_0}_{n_0}.
\end{align}
The computation between the two matrices is paired on the dimension corresponding to $\iota_0$. Note that $\tilde{X}^{\iota_0\iota_1}[(0)()]$ and $\tilde{Y}^{\eta_1\eta_0}[(0)()]$ have the same distribution. The first local computation does the DFT of size $n_1$ followed by the twiddle scaling, while the second local computation does a transposition and the DFT of size $n_0$. An instantiation of the parallel algorithm in the ROTE framework is shown in Figure~\ref{fig:roteC}.

Notice that the besides the advantage of removing redundant communication with the elemental cyclic distribution, the computation of the DFT using ROTE has the property that the {\it{data remains in elemental cyclic distribution after the computation of the 1D DFT}}~\cite{inda2001simple}. In addition, {\it{for a problem size $n = n_0 n_1$, where $p|n_0$ and $p|n_1$, the 1D DFT algorithm requires only one communication stage}} and not three as in the case of the blocked distribution. If one of the conditions is not satisfied then either the mesh is modified or extra communication is required as shown in Inda et. al~\cite{inda2001simple}. We leave the scenario when the above conditions are not satisfied as future work. 

\begin{figure}[t]
    \centering
    \begin{lstlisting}
// create the processing grid
ObjShape gridShape(1);
gridShape[0] = 4;
const Grid g(MPI_COMM_WORLD, gridShape);
    
// create the shape and size of the data tensors
ObjShape shapeX(2);
shapeX[0] = 16;
shapeX[1] = 16;
    
// input and output tensors
DistTensor<complex<double>> A(shapeX, "[(0),()]", g);
DistTensor<complex<double>> B(shapeX, "[(),(0)]", g);
    
fftw_complex *ABuff = (fftw_complex*) A.Buffer();
fftw_complex *BBuff = (fftw_complex*) B.Buffer();
    
// batch DFT_16 in the rows 
fftw_execute_dft(batch_4_dft_16, ABuff, ABuff);
    
// point-wise multiplication with twiddle factors
for(int i = 0; i != 64; ++i) {
    ABuff[i] = ABuff[i] * twidd_local[i];
}

// redistribution
B.AllToAllRedistFrom(A, 0, 0);

// transposition and batch DFT_16 in the rows
fftw_execute_dft(transpose_batch_4_dft_16, BBuff + , BBuff);    
    \end{lstlisting}
    \caption{Code snippet for computing a DFT of size $256$. The 1D DFT is parallelized across $4$ processors on a 1D grid. Computation is done using FFTW. The grid configuration, tensor definition and data distribution is done using ROTE.}
    \label{fig:roteC}
\end{figure}

\section{Experimental Results}

In this section, we describe the experimental setup and the results obtained on the K-Computer, outlining the importance of having a flexible framework that allows one to chose the appropriate algorithm for a given problem size. We begin by presenting the characteristics of the system and then briefly outline some of the implementation details for parallelizing the computation using MPI and OpenMP. We emphasize the simplicity of expressing the computation within the framework. Lastly, we analyze the strong scaling of the 3D DFT for $64^3$, $256^3$ and $1024^3$ using all three decomposition (slab-pencil, pencil-pencil-pencil and volumetric). We give a breakdown of the performance. We show that there is no one decomposition that gives the shortest time to solution.

\subsection{System Configuration}

The Riken AICS K-Computer consists of approximately $80,000$ SPARC$64$ VIIIfx processors. Each compute node has a single CPU with $16$ GB of main memory and a total bandwidth of $64$ GB/s (main memory bandwidth). Each SPARC$64$ VIIIfx has $8$ cores with one thread per core running at $2.0$ Ghz and $6$ MB of L$2$ cache. Each core can do $128$-bit single instruction multiple data (SIMD) instructions, giving a peak performance for double precision fused multiply-add instructions of $128$ GFlops. Each compute node is connected with a 6D mesh/torus network (TOFU interconnect). The TOFU interconnect provides a logical 3D torus network for each job. Each node has $10$ links with a bandwidth of $10 GB/s$ full-duplex and the network allows for multiple pathways to communicate data. The infrastructure allows programs to be adapted to one, two and three dimensional torus networks. The maximum number of processors in one dimension is $54$, while the maximum torus size is $48\times 54\times 32$~\cite{riken}. 

The K-Computer provides its own customize software, ranging from the compilers and the libraries for Fourier transforms and linear algebra operations to the infrastructure for OpenMP and MPI. The current implementation of the DFT framework presented in this work uses FFTW for the local DFT computation and OpenMP to parallelize the computation across the threads. The DFT framework is built on top of ROTE, which allows one to easily describe the data distribution and the communication between the compute nodes. Underneath, ROTE uses MPI to do the collective communication between the compute nodes. The entire framework is compiled with the Fujitsu custom compiler. All optimization are set to \texttt{-O2}. In addition, the \texttt{-fopenmp} and \texttt{-lfftw3} flags are enabled for compilation.

\subsection{OpenMP + MPI Parallelism}

\begin{figure}
    \centering
    \includegraphics[width=0.45\textwidth]{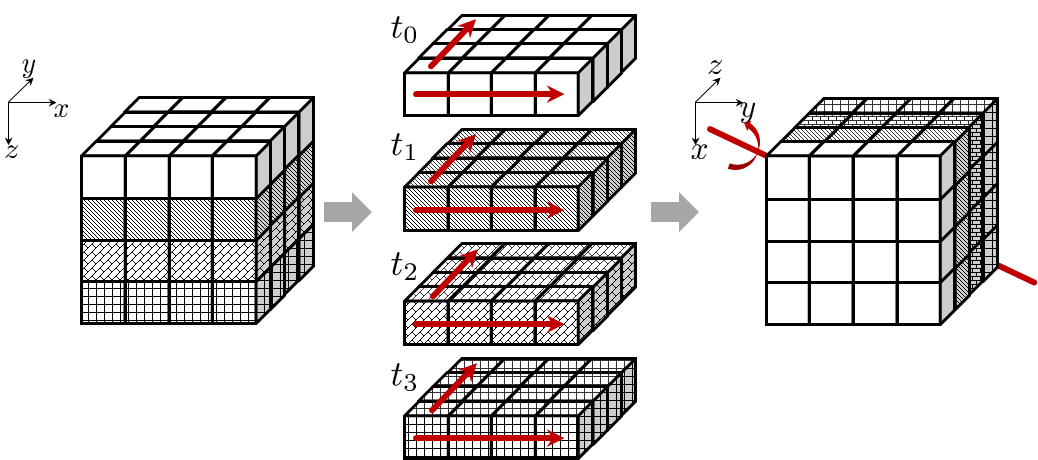}
    \caption{Example of applying thread level parallelism on the local computation using OpenMP. Each processor receives a cube of data points, that is divided between the threads. Each thread takes ownership on that chunk of data and applies its computation and data transformations. In this example, each thread applies computation in two dimensions followed by a rotation across the main diagonal of the data cube.}
    \label{fig:parallel}
\end{figure}

The framework uses OpenMP~\cite{openmp08} parallelism for the local computation and MPI~\cite{MPI2} for the distributed computation. The MPI is hidden away by the ROTE infrastructure. As seen in Figure~\ref{fig:roteC}, where we parallelize the 1D DFT of size $256$ on four processors, ROTE allows users to specify the grid, the tensors, the distribution and the communication. Writing the communication is straightforward. The construction of the processor grid \texttt{g} requires the \texttt{MPI\_COMM\_WORLD} global communicator and the shape of the grip \texttt{gridShape}. Based on the shape of the grid, the global communicator is split accordingly. The construction of the tensors \texttt{A} and \texttt{B} requires the shape of the data tensor, the distribution and the grid on which the data is distributed on. For example, order-2 tensor \texttt{A} is is distributed on the first dimension $[(0)()]$, while order-2 tensor \texttt{B} is distributed on the second dimension $[()(0)]$. Each tensor object provides the necessary functionalities to specify the communication. For the 1D DFT the \texttt{AlltoAllRedistFrom} is used to specify the all-to-all communication described in Figure~\ref{fig:dataredist}.

Parallelizing the computation using OpenMP is made explicit. We uses \texttt{\#pragma omp parallel region} to specify and create a pool of threads. Based on the thread id (\texttt{omp\_get\_thread\_num}), each thread will compute on its own chunk of data as shown in Figure~\ref{fig:parallel}. The figure outlines that once the computation is complete, threads rotate the data. This step is required so that the subsequent MPI all-to-all function call does not incur penalties from accessing data in local memory. We use the clause \texttt{\#pragma omp barrier} to synchronize the threads and the clause \texttt{\#pragma omp master} to specify that only the master thread can perform the communication. Previously, this clause was used around the redistribution function \texttt{AlltoAllRedistFrom} and within the redistribution function other threads were created to perform the packing before and after the MPI call. However, modifications to the original ROTE framework were done and the \texttt{\#pragma omp\_master} was moved around the \texttt{MPI\_Alltoall}, within ROTE's redistribution function. The overall benefit is that a single pool of threads is created up front and those threads are used for both performing the computation and the packing/un-packing of the data. The K-Computer provides eight threads in total, and in the current implementation we use all of them.

\begin{table}
  \centering
  \begin{tabular}{c c c c}
  \toprule
  \# of processors & 1D Grid & 2D Grid & 3D Grid\\
  \midrule
  2 & (2) & - & - \\
  4 & (4) & (2,2) & - \\
  8 & (8) & (4,2) & (2,2,2)\\
  16 & (16) & (4,4) & (4,2,2)\\
  32 & (32) & (8,2) & (4,4,2)\\
  64 & - & (8,8) & (4,4,4)\\
  128 & - & (16,8) & (8,4,4)\\
  256 & - & (16,16) & (8,8,4)\\
  512 & - & (32,16) & (8,8,8)\\
  1k & - & (32,32) & (16,8,8)\\
  2k & - & - & (16,16,8)\\
  4k & - & - & (16,16,16)\\
  8k & - & - & (32,16,16)\\
  16k & - & - & (32,32,16)\\
  32k & - & - & (32,32,32)\\
  \bottomrule
  \end{tabular}
  \newline\newline
  \caption{Table showing the different configurations for the 1D, 2D and 3D grids. The maximum number for each grid represents the maximum size that can be created for the given grid taking into account that the maximum torus size allowed on the K-Computer is $48\times 54\times 32$~\cite{riken}.}\label{table:grid_dims}
  \vspace{-3mm}
\end{table}

\subsection{Results and Discussion} In the following paragraphs, we focus on the performance obtained by our 3D DFT framework on the K-Computer. We compare the execution time of cubic 3D DFTs of size $64^3$, $256^3$ and $1024^3$, using all three algorithms, slab-pencil, pencil-pencil and volumetric decomposition. For all of the configurations, we report strong scaling results, where we keep the problem size fixed and increase the number of processor in each dimension as shown in Table~\ref{table:grid_dims}. We focus on power of two-sized processing grids, where the number of processor does not exceed the number of processor specified in~\cite{riken} in each dimension. We measure the entire 3D DFT computation end-to-end. The actual measurement executes the computation within a loop for $10$ minutes. We then take the average execution time for each experiment. Before the measurement, each MPI process sleeps for approximately one minute. Then $10$ dry runs of the 3D DFT are executed without recording the actual execution time. The two steps are meant to reduce noise on the network. 

{\textbf{A - $64^3$ Problem Size.}} We first present the strong scaling results of the 3D DFT when the problem size is small, i.e. $64^3$ complex data points. As seen in Figure~\ref{fig:1024strongk}, the slab-pencil algorithm for this problem size gives the shortest execution time and it also scales to eight compute nodes (64 cores). For this specific problem size the slab-pencil algorithm cannot be execute on more than eight processors, since we impose the input and output to be in elemental cyclic-ly distributed. Changing the algorithm from slab-pencil to pencil-pencil or volumetric to expose more parallelism, does not provide improvements. Execution suffers slowdowns and in addition the execution of both algorithms tappers off as the number of processors increases. This behaviour is directly related to the all-to-all communication and the number of all-to-alls.

\begin{figure}[t]
    \centering
    \includegraphics[width=0.45\textwidth]{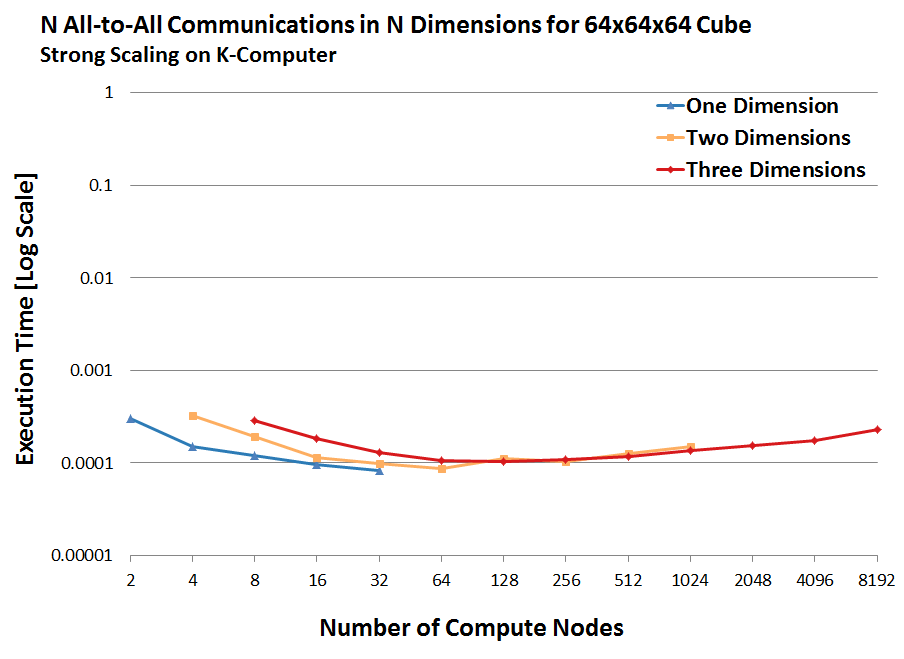}
    \caption{The execution of one, two and three all-to-alls applied on 1D, 2D and 3D grid of processors for the small problem size of $64^3$. Each all-to-all is applied in one dimension. The data cube is distributed between the processors.}
    \label{fig:all2all}
\end{figure}

Depending on the parallel algorithm (slab-pencil, pencil-pencil or volumetric), the 3D DFT may require one or more all-to-all communication steps. Therefore, the scaling behaviour of the 3D DFT is highly dependent on the scaling behaviour of the all-to-all communications. Figure~\ref{fig:all2all} reports the strong scaling of the all-to-all for the same problems size using one/two/three all-to-alls, similar to the three implementations of the 3D DFT. The all-to-alls codes are obtained by removing the local computation and packing. Note that the execution times of the three scenarios exhibit the same trend as the 3D DFT execution times. The slab-pencil approach achieves the shortest execution time and scales to more than eight processors. The pencil-pencil and volumetric variants of the data movement experience slowdowns for a small number of processors and show diminishing returns as the number of processors increased. 

\begin{figure*}[t]
\centering
\minipage{0.5\textwidth}
  \includegraphics[width=\linewidth]{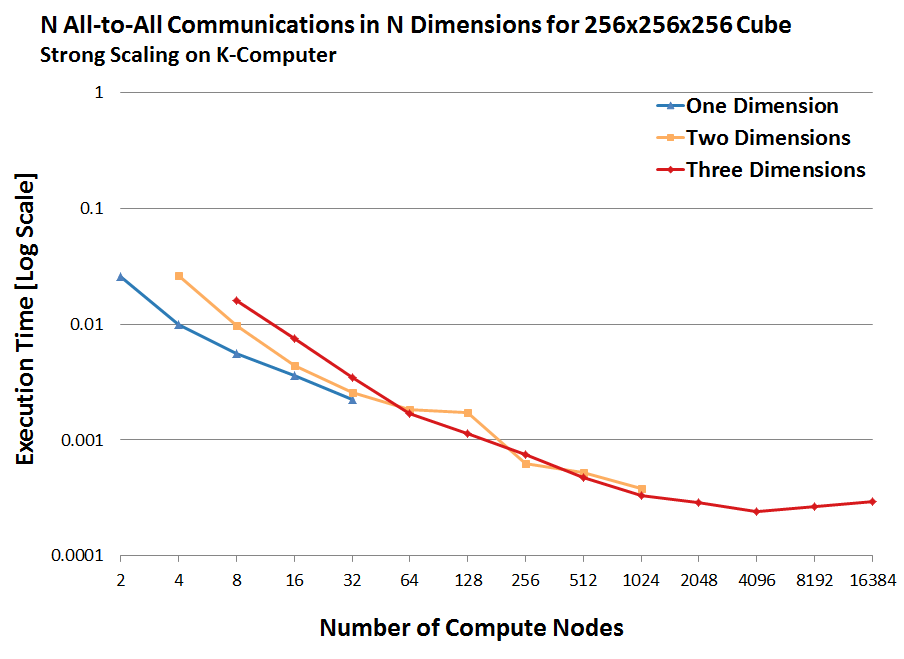}
\endminipage\hfill
\minipage{0.5\textwidth}
  \includegraphics[width=\linewidth]{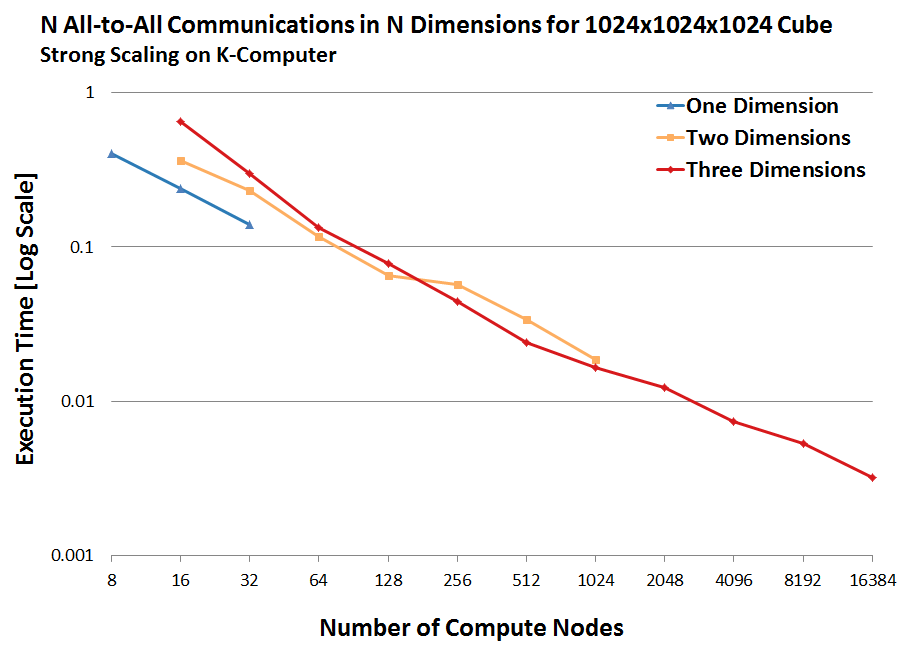}
\endminipage
\caption{The execution of one, two and three all-to-alls applied on 1D, 2D and 3D grid of processors. Each all-to-all is applied in one dimension. The left plot shows the strong scaling of a three dimensional cube of size $256^3$, while the right plot outlines the scaling of a cube of size $1024^3$, respectively. The data cube is distributed between the processors.}
\label{fig:all2all1}
\end{figure*}

The slowdowns and the diminishing returns can be understood by analyzing the lower bounds of the collective communications. Since the K-Computer has its own all-to-all collectives, we discuss the key aspects of the communication by analyzing the lower bounds outlined in Bruck et. al~\cite{bruck1997efficient}. The paper presents the lower bounds for the all-to-all communication as follows
\begin{align}
    \log_{k+1}(p)\alpha + \log_{k+1}(p)\frac{n}{(k+1)p}\beta\label{eq:mst_alg},\\
    \ceil[\big]{\frac{p-1}{k}}\alpha + \ceil[\big]{\frac{p-1}{k}}\frac{n}{p^2}\beta\label{eq:bkt_alg},
\end{align}
where $p$ represents the number of processors, $n$ represents the total amount of data across all $p$ and $k$ represents the number of ports a compute node can send and receive. $\alpha$ represents the start-up cost and $\beta$ represents the inverse of the bandwidth on each of the links. Equation~\ref{eq:mst_alg} represents the lower bound of the all-to-all that minimizes the number of transfers between the nodes, while Equation~\ref{eq:bkt_alg} minimizes the amount of data that is being transferred in each step. The first minimizes the $\alpha$ term, while the second minimizes the $\beta$ term. No one algorithm minimizes both at the same time.

The all-to-all communication becomes latency bound as the number of processors is increased. Not that the latency term $\log_{k+1}(p)\alpha$ ($\ceil[\big]{\frac{p-1}{k}}\alpha$) increases while the bandwidth term $\log_{k+1}(p)\frac{n}{(k+1)p}\beta$ ($\ceil[\big]{\frac{p-1}{k}}\frac{n}{p^2}\beta$) decreases as the number of processors $p$ is increased. Therefore, if $p$ is sufficiently large such that
\begin{align}
    p\geq \frac{n\beta}{(k + 1)\alpha}
\end{align}
for Equation~\ref{eq:mst_alg} and
\begin{align}
    p^2\geq \frac{n\beta}{\alpha}
\end{align}
for Equation~\ref{eq:bkt_alg}, the execution time of the all-to-all becomes latency dominated. In addition, as the number of processors $p$ is increased above those threshold the execution time of the all-to-all should increase. This can be observed in the empirical results outlined in Figure~\ref{fig:all2all}. This provides an explanation to why the DFT for the a size of $64^3$ complex points plateaus as the number of processor is increased. 

The number of ports $k$ lowers the bounds of the all-to-all communication, if the node has sufficient ports/connections to satisfy the $k$ value. The $k$ value provides a possible reason why multiple all-to-alls are slower compared to a single all-to-all for the same number of processors $p$. For example, the $k$ number of ports can be equal to one, two or three when the slab-pencil algorithm is applied on $p = 4$ compute nodes. However, the pencil-pencil algorithm with two all-to-alls on the same amount of processors $p = 4$ has $k$ equal to one. Evaluating the above equations, the slab-pencil algorithm where one all-to-all is used can provide better execution time compared to the other two. However, this effect cannot be seen as $p$ increases, due to the latency aspect that was explained above. While, empirically results as seen in Figure~\ref{fig:all2all} and Figure~\ref{fig:all2all1} seem to back up this aspect, further investigation is needed, especially since the all-to-alls are treated as black boxes. We use the MPI collectives offered by the K-Computer infrastructure.

\begin{figure*}[t]
\minipage{0.5\textwidth}
  \includegraphics[width=\linewidth]{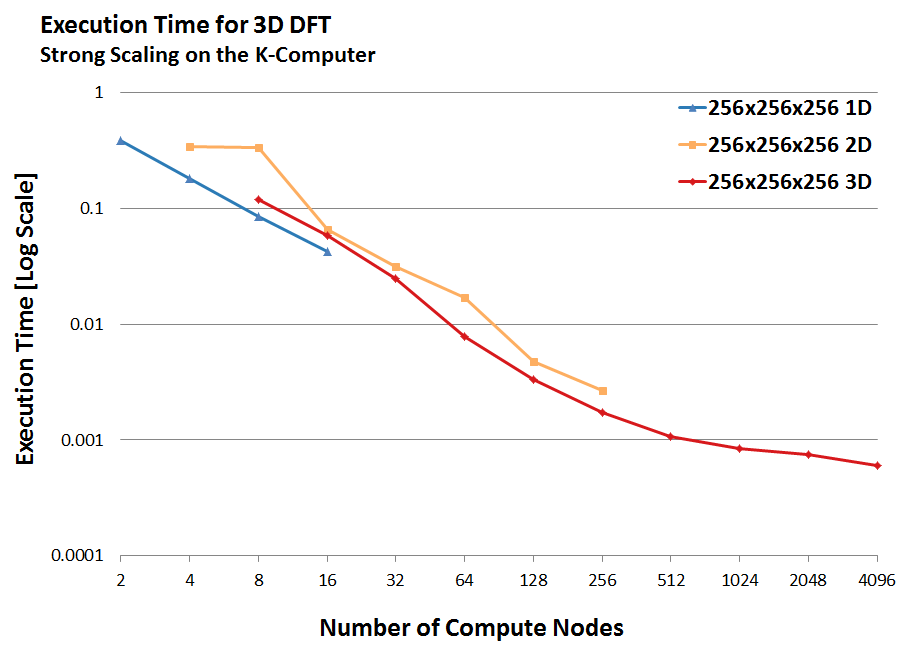}
\endminipage\hfill
\minipage{0.5\textwidth}
  \includegraphics[width=\linewidth]{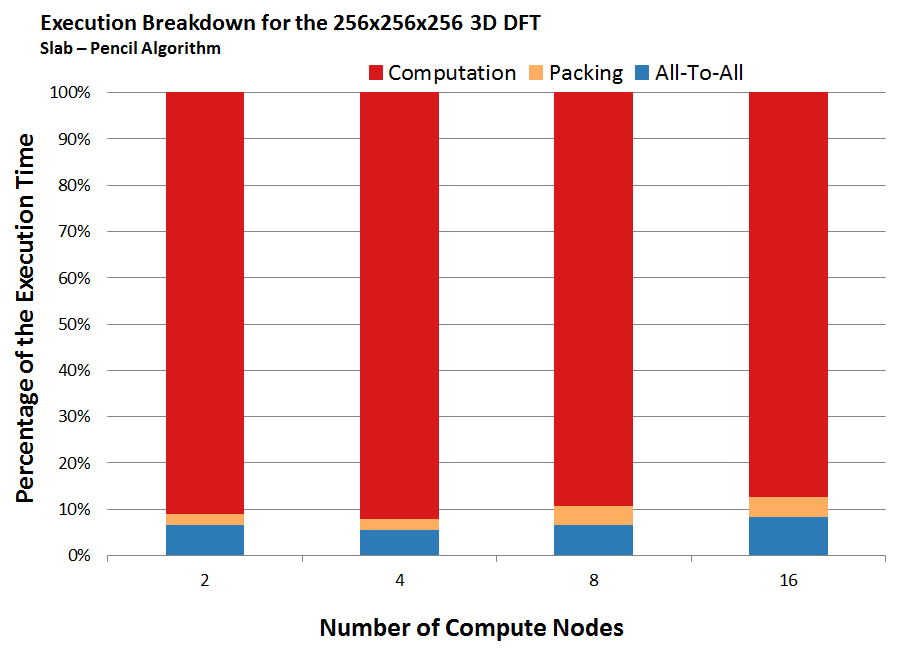}
\endminipage\\
\minipage{0.5\textwidth}%
  \includegraphics[width=\linewidth]{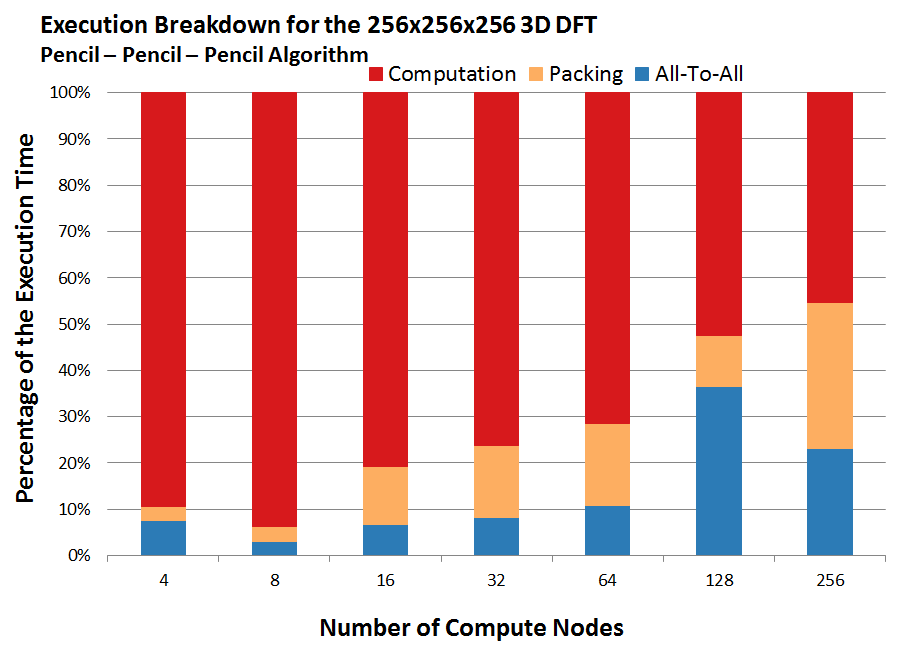}
\endminipage\hfill
\minipage{0.5\textwidth}%
  \includegraphics[width=\linewidth]{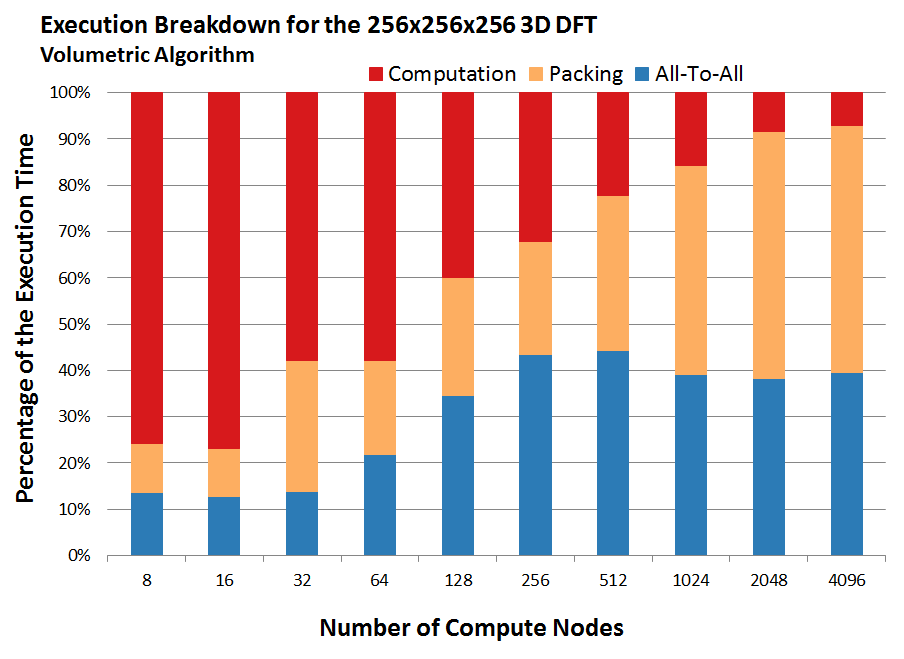}
\endminipage
\caption{Strong scaling results for the 3D DFT of size $256^3$. The top-left plot shows the scaling results for the 3D DFT using all three implementations (blue line for the slab-pencil decomposition, yellow line for the pencil-pencil-pencil decomposition and red line for the volumetric decomposition). The top-right, bottom-left and bottom-right show the execution breakdown for the three decomposition as the number of processors is increased. The blue portion represents the time spent in network communication, while the yellow and red represent the packing and computation times, respectively.}
\label{fig:256_strongk}
\vspace{-2mm}
\end{figure*}

{\textbf{B - $256^3$ 3D DFT.}} In Figure~\ref{fig:256_strongk}, we analyze the strong scaling behaviour of the 3D DFT for a medium size problem of $256^3$ complex data points. Compared to the small size, the amount of data increases by $64$ times, which suggests the problem size can be scaled to larger number of compute nodes. The top left plot outlines the strong scaling of the 3D DFT. Note that when the number of processors is small ($\leq 16$), the slab-pencil decomposition outperforms the other two implementations. However as the number of processors increases, the other two algorithms thrive. Ultimately, the volumetric decomposition offers the best scaling to $4096$ cores/$32768$ threads. Similar behaviour can be seen in the data movement in the left plot in Figure~\ref{fig:all2all1}, where we remove the local computation and packing. Note that as the number of processors is increased, the scaling of the slab-pencil all-to-all diminishes and catches up to the pencil-pencil approach where two all-to-alls are used. Subsequently, the volumetric all-to-all where three all-to-alls are needed catches up and scales to the corresponding number of processors. This emphasizes the need for a more detailed analysis on how the all-to-all is implemented and when to decide to change algorithms for both the all-to-all and the 3D DFT, aspect which we leave for future work.

Data movement across the network is one aspect of the 3D DFT computation, however the local computation and packing also plays a role. The top-right, bottom-left and bottom-right plots in Figure~\ref{fig:256_strongk} show the breakdown of the execution time for all three algorithms, slab-pencil, pencil-pencil and volumetric, respectively. The slab-pencil and pencil-pencil-pencil decomposition show a significant percentage of the execution time to be sent in the local DFT computation. The local computation is basically batches of 1D and 2D DFTs followed by point-wise multiplications that are done outside the DFT. The point-wise computation is not merged since FFTW does not provide such features. For small number of nodes, the amount of local data may reside in local DRAM. As shown in~\cite{Popovici:18}, the DFT is a latency bound computation, therefore applying the double buffering technique on the computation may reduce local computation when the number of processors is small. However, as the number of processors increases the amount of computation is significantly reduced. Note that the data packing becomes a significant part of the overall computation. This suggests that there is still room for improvement by optimizing the packing routines within the ROTE framework or exposing the ROTE packing routines in order to fuse the packing with the computation.

{\textbf{C - $1024^3$ 3D DFT.}} Finally, we report strong scaling results for a large problem size of $1024^3$ complex data points. Figure~\ref{fig:1024strongk} shows the strong scaling behaviour of the entire 3D DFT for this problem size. Similar to the $256^3$ case, the slab-pencil decomposition performs better for small number of processors of up to $32$. However, as the number of processors increases, the volumetric decomposition yields better results. In addition, the volumetric implementation almost linearly scales to $32,768$ compute nodes ($2626,144$ cores), and outperforms for $16,768$ and $32,768$ compute nodes both algorithms presented in~\cite{jung2016parallel}. Note that the pencil-pencil decomposition follows the volumetric decomposition until scaling the problem to $512$ compute nodes. Similar to the $256^3$ case, computation can still be improved for the case where the number of processors is small. The same cannot be said as the number of processors is increased beyond $4096$ compute nodes. Computation and packing becomes insignificant compare to the data movement. The execution of the 3D DFT for large number of nodes follows within $10\%$ the execution of the all-to-all communication depicted in Figure~\ref{fig:all2all1}.

\section{Conclusion}
In this paper, we presented a flexible framework for implementing parallel multi-dimensional DFTs on multi-dimensional processing grids. Specifically, we show that for different input problem size and different computational resource availability, different parallel multi-dimensional DFT algorithms are necessary for attaining efficient performance. In addition, we show that it is necessary to parallelize within the 1D DFT in order to scale the computation of the multi-dimensional DFT towards higher number of processing units. Despite incurring more rounds of communication, we show that for large enough data sizes, parallelizing with the 1D DFTs can improve the overall execution time over the conventional approach of simply parallelizing across multiple 1D DFTs.

While we showed improved performance as we scale to larger number of nodes, further improvements to performance can be attained. As shown in Figure~\ref{fig:256_strongk}, a drawback of the current ROTE framework is that a local packing step is required to repack the data back into elemental-cyclic form after the collective communication. The time for the repacking is significant as it could be as much as 50\% of the overall execution time. One possibility for reducing this packing time is to expose the packing performed by ROTE or allow ROTE to accept data in packed form. This is something we are currently exploring. We also believe that the ROTE notation for describing data layout and processing grid can be extended to other architectures such as GPUs and multi-GPU systems. This can be achieved by replacing the MPI routines with the appropriate data movement primitives for GPUs and multi-GPUs. As the hierarchy of threads, thread blocks, streaming processors, and GPUs can be described as a multi-dimensional array, this suggest that ROTE can be used as the notation for porting the existing code to GPUs and multi-GPUs systems.

\begin{acks}
The authors would like to thank Dr. Imamura Toshiyuki and the Advanced Institute for Computational Science at RIKEN for granting access to the K-computer. This work was sponsored partly by the DARPA PERFECT and BRASS programs under agreements HR0011-13-2-0007 and FA8750-16-2-003, and NSF through award ACI 1550486. The content, views and conclusions presented in this document are those of the author(s) and do not necessarily reflect the position or the policy of the sponsoring agencies.
\end{acks}

\bibliographystyle{unsrt}
\bibliographystyle{ACM-Reference-Format}

\end{document}